\documentclass[letterpaper,aps,prl,twocolumn,floats,floatfix,amssymb,amsfonts,amsmath]{revtex4}

\usepackage{pgf}
\usepackage[utf8x]{inputenc}

\renewcommand{\vec}[1]{{\mathbf{#1}}}
\newcommand{\dd}[0]{\ensuremath{\mathrm{d}}}

\begin{document}

\title{Solution to the Cosmic Ray Anisotropy Problem}

\author{Philipp Mertsch}
\author{Stefan Funk}
\affiliation{Kavli Institute for Particle Astrophysics \& Cosmology, 2575 Sand Hill Road, M/S 29, Menlo Park, CA 94025, USA}

\begin{abstract}
In the standard diffusive picture for transport of cosmic rays (CRs), a gradient in the CR density induces a typically small, dipolar anisotropy in their arrival directions. This has been widely advertised as a tool for finding nearby sources. However, the predicted dipole amplitude at TeV and PeV energies exceeds the measured one by almost two orders of magnitude. Here, we critically examine the validity of this prediction which is based on averaging over an ensemble of turbulent magnetic fields. We focus (1) on the deviations of the dipole in a particular random realisation from the ensemble average and (2) the possibility of a misalignment between the regular magnetic field and the CR gradient. We find that if the field direction and the gradient direction are close to $\sim 90^\circ$, the dipole amplitude is considerably suppressed and can be reconciled with observations, which sheds light on a long-standing problem. Furthermore, we show that the dipole direction in general does \emph{not} coincide with the gradient direction, thus hampering the search for nearby sources.
\end{abstract}

\maketitle

Cosmic rays (CRs) with energies between hundreds of MeV and at least a few PeV are commonly believed to be of galactic origin.  In the standard picture, the high degree of isotropy in their arrival directions is interpreted as evidence for diffusion as providing the necessary mechanism for efficiently randomising their directions. On the other hand, in the case of a not perfectly symmetric distribution of sources with respect to the observer, a small degree of anisotropy, to first order a dipole in the arrival direction of cosmic rays, is to be expected. In particular, a (few) nearby source(s) can have a dominant effect on the distribution of arrival directions which is why observation of a dipole anisotropy has been advertised as a means of discovering these nearby sources~\cite{1995ICRC....3...56P,Buesching:2008hr,Sveshnikova:2013ui}. Lately, this idea has gained currency in the context of finding the necessarily nearby (because of cooling losses) source(s) of high-energy electrons and positrons~\cite{DiBernardo:2010is,Borriello:2010qh,Linden:2013mqa} which is/are causing the rise in the positron fraction~\cite{Adriani:2008zr,FermiLAT:2011ab,Aguilar:2013qda}.

Given the high degree of isotropy, a perturbative approach is adopted in CR transport models, expanding the phase space density $f(\vec{r}, \vec{p}, t)$ into an isotropic part $f_0(\vec{r}, p, t)$ and a small correction, $f_1(\vec{r}, \vec{p}, t)$. $f_1(\vec{r}, \vec{p}, t)$ is then related to the gradients of $f_0(\vec{r}, |\vec{p}|, t)$, the momentum gradient leading to the well-known Compton-Getting effect~\cite{1935PhRv...47..817C}; here, we focus on the spatial gradient. In a simple model of isotropic diffusion, the amplitude $a$ of the dipole anisotropy, the relative difference between the fluxes in the maximum and minimum directions, $\phi_\text{max}$ and $\phi_\text{min}$, computes as~\cite{Ginzburg:1990sk}
\begin{equation}
a = \frac{\phi_\text{max} - \phi_\text{min}}{\phi_\text{max} + \phi_\text{min}} = \frac{3 D}{v} \frac{| \nabla f_0 |}{f_0} \, ,
\label{eqn:aIso}
\end{equation}
where $D$ is the (local) spatial diffusion coefficient and $v \approx c$ is the CR speed. The dipole direction is opposite to that of the CR gradient. For a given distribution of sources and extrapolating the diffusion coefficients measured through secondary-to-primary ratios like B/C at $\text{GV}$ to $\text{TV}$ rigidities, one can first compute the CR density $f_0$ and through eq.~\ref{eqn:aIso} the dipole amplitude. The rigidity-dependence of the dipole amplitude results from both $D$ and $| \nabla f_0 | / f_0$.

\begin{figure}[b]
\input{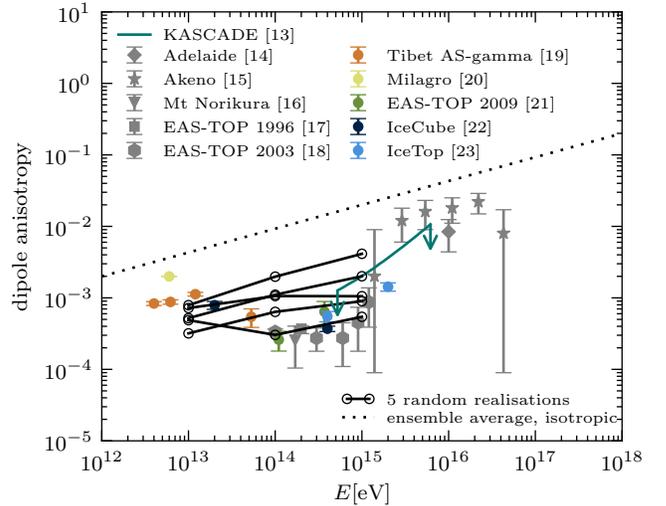}
\caption{The dipole anisotropy in the arrival directions of CRs, as predicted by an isotropic diffusion model~\cite{Blasi:2011fm} (dotted line) and measured by a variety of experiments~\cite{Antoni:2003jm,Gerhardy:1984wq,Kifune:1985vq,Nagashima:1990ze,Aglietta:1996sz,Aglietta:2003uc,Amenomori:2005dy,Abdo:2008aw,Aglietta:2009mu,Abbasi:2011zka,Aartsen:2012ma}.
The black filled circles, connected by solid lines, mark the dipole anisotropy predicted in five random realisations of the turbulent magnetic field and assuming a misalignment between background magnetic field and CR gradient close to $90^\circ$.}
\label{fig1}
\end{figure}

Over the last decades, a large set of measurements of the dipole anisotropy has been accumulated, at energies above a few TeV mostly from extensive airshower arrays~\cite{Antoni:2003jm,Gerhardy:1984wq,Kifune:1985vq,Nagashima:1990ze,Aglietta:1996sz,Aglietta:2003uc,Amenomori:2005dy,Abdo:2008aw,Aglietta:2009mu,Abbasi:2011zka,Aartsen:2012ma}. The dipole amplitude decreases from $\sim 10^{-3}$ at $10 \, \text{TeV}$ to $\sim 10^{-4}$ between $100 \, \text{TeV}$ and $1 \, \text{PeV}$ before it increases again. Here we limit ourselves to energies below a few PeV where CRs are certainly of galactic origin and where the composition is predominantly p and He. We show these measurements together with the prediction from a simple diffusion model in Fig.~\ref{fig1}. It is evident (as has been known for a while~\cite{2005JPhG...31R..95H,2006AdSpR..37.1909P}) that the diffusion model overpredicts the dipole amplitude by almost two orders of magnitude around $1 \, \text{PeV}$.

This apparent discrepancy, dubbed the `CR anisotropy problem', has led to various modelling attempts. For example, it was pointed out that the point- and transient-like nature of CR sources, like supernova remnants (SNRs), leads to fluctuations for different positions and periods of observations~\cite{2006AdSpR..37.1909P}; however, it has been shown by Monte Carlo methods~\cite{Blasi:2011fm,Pohl:2012xs} that even under favourable conditions an observation position or period in agreement with the measured dipole amplitude is unlikely. Furthermore, it has been suggested that the CR gradient would be smaller than usually predicted if the diffusion coefficient was allowed to spatially correlate with the sources of turbulence in the interstellar medium~\cite{Evoli:2012ha}, e.g. SNRs, instead of being assumed to be constant. However, while this can alleviate some of the tension at hundreds of GeV, the predicted anisotropy at hundreds of TeV is still more than an order of magnitude too large. 

It is instructive to revisit the derivation of eq.~(\ref{eqn:aIso}) to investigate which assumptions need to be relaxed in order to reconcile the predicted dipole amplitude with measurements. For this, we adopt the framework of quasi-linear theory~\cite{1966ApJ...146..480J} in which the magnetic field $\vec{B} = \vec{B}_0 + \delta \vec{B}$ is the sum of a regular field $\vec{B}_0$ and a turbulent field $\delta \vec{B}$ with $\delta B^2 = \int \dd^3 k \, \delta\vec{B}^2(k) \ll |\vec{B}_0 |^2$, and the turbulent field is evaluated along unperturbed trajectories. 
(The dipole anisotropy can also be computed in a more general framework, see e.g., Ref.~\cite{Ahlers:2013ima}, however, at the expense of not predicting the scattering rate from first principles.)
Without loss of generality, we take $\vec{B}_0$ to point into the $x$-direction which is also defining the pitch angle $\mu = p_x / p$ of a particle of momentum $\vec{p}=(p_x, p_y, p_z)^T$. Under additional assumptions (e.g. after gyro-phase averaging, see, e.g.,~\cite{Ginzburg:1990sk} for details) the \emph{ensemble averaged} distribution function can be expanded into an isotropic part and a pitch-angle dependent part: $\bar{f}(\vec{r}, \vec{p}, t) = f_0(\vec{r}, p, t) + f_1(\vec{r}, p, \mu, t)$. The first moment of the anisotropic part then determines the dipole amplitude (along the regular field),
\begin{equation}
a = \frac{\frac{3}{2} \int_{-1}^{1} \dd \mu \, \mu \, f_1(\mu)}{f_0} = \frac{3}{v} \frac{|\partial f_0 / \partial x|}{f_0} D_{\parallel} \, .
\label{eqn:aQLT}
\end{equation}
Here, $D_{\parallel} = D_{xx}$ is the parallel diffusion coefficient.
%Note that contributions due to perpendicular diffusion~\cite{2007ApJ...661..185S} are of higher order in $V_A/c$ where $V_A$ is the phase speed of the turbulent modes.

The derivation of eq.~(\ref{eqn:aQLT}) reveals two important limitations: First, the amplitude of the dipole anisotropy is an ensemble average, much like the underlying distribution function $F$ is the average for the gaussian random field $\delta\vec{B}$. For propagation on galactic scales, this is commonly justified by assuming ergodicity: For propagation times $t \gg L^2/D$ where $L$ is the outer scale of turbulence in the ISM, $L \sim 100 \, \text{pc}$, CRs from sources at kpc distances will experience many different field configurations before observation. However, what is observed is not the time-averaged distribution function, but just a snapshot which, as we will see, is affected by the (local) realisation of $\delta\vec{B}$. (Also, note that typical observation times are shorter than the coherence time of the magnetic field turbulence.) The effect of the local field configuration was recently considered in the context of observed small scale anisotropies~\cite{Ahlers:2013ima,Giacinti:2011mz}. In particular, it was shown that the local $\delta\vec{B}$ is leading to non-diffusive behaviour, dynamically generating and destroying correlations on all angular scales within a few scattering times $\tau_{\text{sc}}$~\cite{Ahlers:2013ima}.

Second, eq.~(\ref{eqn:aQLT}) stresses the anisotropic nature of diffusion in the ISM and the possibility of misalignment between CR gradient $\nabla f_0$ and regular field $\vec{B}_0$. For the case of a perfect \emph{misalignment} of $\nabla f_0$ and $\vec{B}_0$, the amplitude of the ensemble averaged dipole vanishes. In the absence of deviations from the ensemble average, this would already solve the CR anisotropy problem.

To investigate the interplay between these two effects, i.e. the deviation from the ensemble average in specific (local) realisations of $\delta\vec{B}$ on the one hand and the possible misalignment between $\nabla f_0$ and $\vec{B}_0$ on the other hand, we employ a numerical simulation of the transport of charged particles through a turbulent magnetic field. Specifically, we backtrack particles, i.e. we follow particles of opposite charge backwards in time, injecting them in the opposite direction from which they would be observed. For observed directions $\vec{n_i} = {\vec{p}_i(0)}/p$ and positions ${\vec{r}_i(0)} = 0$ we obtain a set of back-tracked trajectories, ${\{ ( \vec{r}_i(t)}, \vec{p}_i(t) ) \}$. With these, and due to Liouville's theorem, we can compute the flux at $t = 0$, $\vec{r} = 0$ in direction $\vec{n}_i$ from a given phase space density at an earlier time $-T < 0$, $f(\vec{r}_i(0), \vec{p}_i(0)) = f(\vec{r}_i(-T), \vec{p}_i(-T))$.

The relativistic equations of motion for charged particles are solved with an 5th order adaptive Runge-Kutta algorithm~\cite{Sutherland:2010vx} and we consider particles with rigidities of $10$, $100$ and $1000 \, \text{TV}$, neglecting the finite energy resolution of the experimental data. The level of turbulence $\eta \equiv \delta B^2 / (\delta B^2 + B_0^2)$ is very uncertain, especially in the local ISM. On large scales, Faraday rotation measurements point to regular fields as low as $B_0 \simeq (1.4 \pm 0.2) \, \mu\text{G}$ which with a total field of $6 \, \mu\text{G}$ (from equipartition) gives $\eta \simeq 0.96$~\cite{2013pss5.book..641B}. On the other hand, to reproduce the grammage inferred from nuclear secondary-to-primary ratios, a much smaller $\eta \lesssim 0.02$ would be needed~\cite{DeMarco:2007eh}. To bracket these vastly different estimates (and to ease comparison with other numerical studies~\cite{Casse:2001be,Giacinti:2011ww}) we adopt two values, $\eta = 1$ and $0.1$. The turbulent field is computed on a set of nested grids~\cite{Giacinti:2011ww} with a Kolmogorov spectrum, an outer scale $L = 100 \, \text{pc}$ and a total (RMS) field strength $4 \,\mu \text{G}$. We have checked that averaging over many different realisations of $\delta \vec{B}$, we recover diffuse behaviour. In particular, adopting the parameters of Refs.~\cite{Casse:2001be,Giacinti:2011ww} we reproduce the inferred pitch-angle scattering times and diffusion coefficients.

\begin{figure*}[!tbh]
	\centering
	\begin{minipage}[t]{\columnwidth}
		\input{map_p_KOLa_1000TeV_1000yr_nside64_it100_s01-50_ang_90.pgf}
		\caption{Sky map of dipole directions in 50 random realisations of the local turbulent magnetic field ($\eta=1$) at $1 \, \text{PV}$. The centre and radius of each black circle shows the dipole direction and amplitude in one random realisation, respectively. The yellow star shows the direction of the assumed CR gradient, the green diamond the predicted value from an isotropic diffusion model and the red square the average of the 50 magnetic field configurations.}
		\label{fig:map_iso}
	\end{minipage}
	\hfill
	\begin{minipage}[t]{\columnwidth}
		\input{hist2d_p_KOLa_1000TeV_1000yr_nside64_it100_s01-50.pgf}
		\caption{The distribution of dipole amplitudes as a function of the longitude of the CR gradient at $1 \, \text{PV}$ for $\eta=1$. Each vertical slice is the normalised histogram for a gradient direction. We also show the median (orange dashed line), and amplitude of the (vectorial) mean (red solid line), together with the prediction for isotropic diffusion (green dashed line). The cyan solid line and grey band show the KASCADE upper limit and EAS-TOP measurement at $\sim 1 \, \text{PeV}$, respectively.}
		\label{fig:hist_iso}
	\end{minipage}
	\begin{minipage}[t]{\columnwidth}
		\input{map_p_rKOLa01_1000TeV_10000yr_nside64_it100_s01-50_ang_90.pgf}
		\caption{Same as Fig.~\ref{fig:map_iso}, but for a small turbulent field on top of a regular field ($\eta = 0.1$), indicated by the blue cross.}
		\label{fig:map_QLT}
	\end{minipage}
	\hfill
	\begin{minipage}[t]{\columnwidth}
		\input{hist2d_p_rKOLa01_1000TeV_10000yr_nside64_it100_s01-50.pgf}
		\caption{Same as Fig.~\ref{fig:hist_iso}, but for a small turbulent field on top of a regular field ($\eta = 0.1$).}
		\label{fig:hist_QLT}
	\end{minipage}
\end{figure*}

We have confirmed that our computation reproduces the expected anisotropies at time $0$ for different combinations of (in)homogeneous and (an)isotropic distribution functions at time $-T$. Trivially, a homogeneous and isotropic phase space density at $-T$ leads to no anisotropy. A homogeneous distribution of dipoles at $-T$ leads not only to a dipole but also to power at smaller multipole moments $\ell$, all of which are eventually decaying exponentially with a time constant $\tau_{\text{sc}} / \ell(\ell+1)$~\cite{Ahlers:2013ima}. The scenario we are most interested in here is an initial gradient in the phase space distribution: After a few $\tau_{\text{sc}}$, the distribution of arrival directions converges, irrespective of the initial angular distribution. We observe anisotropies extending to the highest multipoles allowed by our angular resolution which are eventually all powered by the spatial gradient in the initial distribution function.

Every spatial distribution at times $-T$ can be expanded into a spatially homogeneous part, a gradient and higher derivatives. We assume that the higher derivatives are subdominant and adopt the (ensemble averaged) gradient from the diffusion model. We read off this gradient for the average source distribution from Fig.~2a of Ref.~\cite{Blasi:2011fm} adopting their parametrisation of the diffusion coefficient measured from B/C.

We start by presenting our results for the case of isotropic turbulence without a regular field, i.e. $\eta = 1$. In Fig.~\ref{fig:map_iso}, we show the dipole directions by the black circles, obtained for 50 random realisations of the magnetic field, a CR gradient in $(long, lat) = (90^\circ, 0^\circ)$ and $ 1\, \text{PV}$ particles. There is considerable scatter in the dipole \emph{directions} around the predicted value from the diffusion model (indicated by the green diamond). The mean dipole of the 50 random realisations (indicated by the red square), however, reproduces the predicted dipole very well. There is also scatter in the \emph{amplitudes} of individual dipoles, and on average their amplitude is larger than that of the predicted dipole. This is also shown in Fig.~\ref{fig:hist_iso}, where we have varied the direction of the CR gradient: Each vertical slice is a (normalised) histogram of the distribution of the dipole amplitudes. The red (green) line marks the average amplitude from the 50 random realisations (the predicted amplitude from the isotropic diffusion model), corresponding to the sizes of the red square (green diamond) in Fig.~\ref{fig:map_iso}. As expected there is no preferred direction. At the same time, while there is some scatter around the predicted dipole amplitude, it is too little to explain the small experimental upper limit and measurement of a few times $10^{-4}$.

The results for the case with a strong regular field differ significantly. Although $\nabla f_0$ is still at $(long, lat) = (90^\circ, 0^\circ)$, the dipole directions now cluster around the $\vec B_0$ direction which is at $(long, lat) = (0^\circ, 0^\circ)$, see Fig.~\ref{fig:map_QLT}. It is also apparent that for this misalignment the amplitudes are markedly suppressed. While this was already expected for the ensemble averaged dipole, see eq.~\ref{eqn:aQLT}, there is considerable scatter, both in amplitude and in direction. We emphasise that this is due to the random nature of $\delta\vec{B}$ which is not accounted for in eq.~\ref{eqn:aQLT}, and that the dipole is likely in the direction of the \emph{total} $\vec{B}$ as sampled by the CR trajectories. In Fig.~\ref{fig:hist_QLT}, we show the distribution of dipole amplitudes as a function of longitude of $\nabla f_0$ (here, that is the angle between $\nabla f_0$ and $\vec B_0$). The distribution shows the expected cosine-behaviour and for most angles the scatter is small. For near-to-perfect misalignment, however, we find that the majority of random realisations has a dipole amplitude below the KASCADE upper limit and of order $20 \, \%$ are consistent with the EAS-TOP 2003 measurement.

In Fig.~\ref{fig1}, we have shown the dipole amplitudes for five random field realisations at $10, 100 \, \text{TeV}$ and $1 \, \text{PeV}$ (assuming protons) for an angle $\sim 90^\circ$ between CR gradient and regular field. We note that while for three random realisations the amplitudes increase with energy as expected, the other two show non-trivial energy dependences, one even closely tracing the experimentally observed decline and rise from $10$ to $100 \, \text{TeV}$ and on to $1 \, \text{PeV}$. This non-standard energy dependence is evidence of particles of different rigidities sampling different parts of the local turbulent field.

One might wonder how a possible energy dependence of the gradient direction would affect our results. First, we note that to achieve the necessary suppression, the CR gradient and the regular magnetic field need to be perpendicular only to within $\sim 10^{\circ}$ (see Fig.~\ref{fig:hist_QLT}). Second, it has been shown in computations of the CR gradient, that its direction is relatively stable, i.e. it varies by less than $10^{\circ}$ for a relatively wide range of energies~\cite{Blasi:2011fm}. Therefore, even with finite energy resolution the suppression of the dipole can be maintained for energies between $\sim 10^{14}$ and $10^{15} \, \text{eV}$ whereas outside this interval the misalignment could be less severe, thus causing less suppression of the dipole amplitude.

To a certain degree, the rigidity-dependent sampling of the magnetic field could also cause the rise in anisotropy above a few PeV which seems to be observed in earlier measurements~\cite{Gerhardy:1984wq,Kifune:1985vq}. (We caution however that contemporary experiments, e.g.~\cite{Antoni:2003jm}, have so far only placed upper limits.) After a few scattering times, particles of rigidity $\mathcal{R}_{\text{Pev}} \, \text{PeV}$ cover on average distances of a few times $\sqrt{4 D_{\parallel} \tau_{\text{sc}}} \simeq 100^{q-1} \, \text{pc} \, (\delta B/  B_0)^{-2} ( \mathcal{R}_\text{PV} / B_{0, \mu\text{G}} )^{2-q}$ (here, $q = 5/3$ is the Kolmogorov spectral index). With the parameters adopted above for the $\eta = 0.1$ case, this computes as a few hundred parsecs at $1 \, \text{PV}$ and more than doubles with every decade in rigidity. Therefore, at higher rigidities a larger part of the turbulent field is being sampled, with the large-scale modes ($1/k \ll r_L$) effectively contributing to the background field. This already leads to the non-standard rigidity-dependence for the energy range shown in Fig.~\ref{fig1}, but as the amplitude of the turbulent modes grows with scale, this effect is expected to get stronger with rigidity. Furthermore, in numerically computing the trajectories of CRs, we have neglected the possibility of escape from the CR halo which is only valid if the distances travelled are much smaller than the size of the CR halo. Beyond PV rigidities, this assumption breaks down which will lead to a higher anisotropy amplitude. Finally, at even higher rigidities, the gyroradius approaches the scale of variations in $\vec{B}_0$, such that gradient and curvature drifts cannot be neglected anymore, leading to additional contributions to the anisotropy. We leave a detailed study of these effects to future work.

Our conclusions are two-fold: First, it is not possible to determine the direction of the CR gradient (and thus the direction of the closest, dominant source or the bulk of sources) from the dipole direction: For strong turbulence, there is considerable scatter between  the CR gradient and the dipole directions in random realisations; for weak turbulence, the dipole directions scatter around the regular field direction rather than around the CR gradient. Second, the small observed dipole amplitudes between $\sim 100 \, \text{TeV}$ and $1 \, \text{PeV}$ can be understood in the presence of a strong regular field only if the CR gradient and the regular field are almost maximally misaligned. This opens up the possibility of determining the direction of the regular magnetic field, if the CR gradient could be reliably predicted or determined by other means, e.g. diffuse gamma-ray backgrounds~\cite{collaboration:2009ag}. We conclude by noting that in a similar way that the dipole depends on the regular field direction, the higher multipoles of the arrival direction encode information about the higher moments of the (turbulent) local magnetic field. While it is computationally very challenging to solve the inverse problem of inferring the latter from the former, observations of CR arrival directions might soon prove one of the most valuable probes of the nearby galactic magnetic field.

It is a pleasure to thank Markus Ahlers for stimulating discussions. PM is supported by DoE contract DE-AC02-76SF00515 and a KIPAC Kavli Fellowship. We acknowledge use of the \texttt{HEALPix} package~\citep{Gorski:2004by}.

%------------------------------------------------------------------------------
%------------------------------------------------------------------------------

\end{document}